\documentstyle[preprint,aps,prd,epsfig]{revtex}
\begin{document}  
\draft
\preprint{OCIP/C 95-12, hep-ph/9510245}
\title{PROPERTIES OF THE STRANGE AXIAL MESONS \\
IN THE RELATIVIZED QUARK MODEL}
\author{Harry G. Blundell, Stephen Godfrey\footnote{e-mail: 
godfrey@physics.carleton.ca} and Brian Phelps}
\address{Ottawa-Carleton Institute for Physics \\
Department of Physics, Carleton University, Ottawa CANADA, K1S 5B6}
\date{September 1995}
\maketitle
\begin{abstract}
We studied properties of the strange axial mesons in the relativized 
quark model.  We calculated the $K_1$ decay constant in the quark 
model and showed how it can be used to extract 
the $K_1 (^3P_1) - K_1 (^1P_1)$ 
mixing angle ($\theta_K$) from the weak decay $\tau \to K_1 \nu_\tau$. The 
ratio 
$BR(\tau \to \nu_\tau K_1 (1270))/BR(\tau\to \nu_\tau K_1(1400))$ is 
the most sensitive measurement and also the most reliable since the 
largest of the theoretical uncertainties factor out. However the 
current bounds extracted from the TPC/Two-Gamma collaboration measurements 
are rather weak:  we typically obtain $-30^o \lesssim \theta_K \lesssim 50^o$ at 
68\% C.L.
We also calculated the strong OZI-allowed  
decays in the pseudoscalar emission model and the flux-tube breaking 
model and extracted a $^3P_1 - ^1P_1$ mixing angle of 
$\theta_K \simeq 45^o$.  
Our analysis also indicates that the heavy quark limit does not give 
a good description of the strange mesons.
\end{abstract}

\section{INTRODUCTION}
\label{sec:intro}

The strange axial mesons offer interesting possibilities for the 
study of QCD in the non-perturbative regime through the mixing of 
the $^3P_1$ and $^1P_1$ states.  In the SU(3) limit these states
do not mix, just as the $a_1$ and $b_1$ mesons do not mix.  
For a strange quark mass greater than the
up and down quark masses SU(3) is broken 
so that the $^3P_1$ and $^1 P_1$ states mix to
give the physical $K_1$ states.  In the heavy quark limit where the strange
quark becomes infinitely heavy, the light quark's spin couples with the
orbital angular momentum resulting in the light quark having 
total angular momentum  $j={3 \over 2}$ in one state and
$j={1 \over 2}$ in the other state, each state having distinct 
properties\cite{derujula,rosner86,godfrey91}. By studying the
strange axial mesons and comparing them to the heavy quark limit one
might gain some insights about hadronic properties in the {\it soft} QCD
regime. 

Recently, the TPC/Two-Gamma Collaboration has presented measurements for
the decays $\tau^- \to \nu_\tau K_1^- (1270)$ and 
$\tau^- \to \nu_\tau K_1^- (1400)$ \cite{tpc95}.  It is expected that the
LEP, CLEO, and BES collaborations,  
with their large samples of $\tau$'s, will be able to
study these decays in further detail \cite{towers}. These decays provide
another means of studying  
$^3P_1 -^1 P_1$ mixing of the strange axial mesons 
in addition to using their partial decay widths and masses.

In this paper we study the properties of the strange axial mesons in the
context of the relativized quark model \cite{godfrey85,godfrey85b}.  
We compare the experimental measurements to the predictions of the model 
to extract the   $^3P_1 -^1 P_1$ mixing angle ($\theta_K$).  
Comparing both the
experimental measurements and model results to various limits
helps in understanding the nature of QCD in the {\it soft} regime.

We begin in Sec. II with a brief description of the relativized quark
model and a description of the $^3P_1 -^1 P_1$ mixing.  By comparing the
mass predictions of the quark model to the observed 
$K_1$ masses we obtain our first
estimate for $\theta_K$.  In Sec. III we calculate the $K_1$ decay
constants using the mock-meson approach and use the results to obtain a
second estimate of $\theta_K$.  In Sec. IV we study the strong
decay properties
of these states using the pseudoscalar emission model\cite{godfrey85}
and the flux-tube breaking model
\cite{kokoski} and use the results as another way of measuring the 
$^3P_1 - ^1P_1$ mixing angle.  
When appropriate we examine the non-relativistic and
heavy quark limits to gain insights into the underlying dynamics.  
Various aspects of 
the phenomenology of the strange axial mesons have also been studied by
Suzuki in a series of recent papers \cite{suzuki93,suzuki94}
using approaches complementary to ours.

\section{THE $K_1$ MASSES and $^3P_1 -^1P_1$ MIXING}

In this section we give a very brief description of the relativized
quark model \cite{godfrey85,godfrey85b}.  
The spin-orbit contributions in particular will be
important in understanding the $^3P_1 -^1 P_1$ mixing.  The model 
is not derived from first principles but rather is motivated by 
expected relativistic properties.  Although progress is being made 
using more rigorous approaches, the relativized quark model describes 
the properties of hadrons reasonably well and presents an approach 
which can give insights into the underlying dynamics that can be 
obscured in the more rigorous approaches.  

The basic equation of the model is the rest
frame Schr\"{o}dinger-type equation.
The effective potential, $V_{q\bar{q}} (\vec{p},\vec{r})$,
is described by a Lorentz-vector one-gluon-exchange interaction at 
short distances and a Lorentz-scalar linear confining interaction.  
$V_{q\bar{q}} (\vec{p},\vec{r})$
was found by equating the scattering amplitude of free quarks,
using a scattering kernel with the desired Dirac structure, with the effects
between bound quarks inside a hadron \cite{gromes}.
Due to relativistic effects the potential is
momentum dependent in addition to being co-ordinate dependent.
The details of the model can be found in Ref. \cite{godfrey85}.
To first order in $(v/c)^2$, $V_{q\bar{q}} (\vec{p},\vec{r})$
reduces to the standard non-relativistic result:
\begin{equation}
V_{q\bar{q}} (\vec{p},\vec{r}) \to V(\vec{r}) =
H^{conf}_{q\bar{q}}  +H^{cont}_{q\bar{q}} + H^{ten}_{q\bar{q}}
+ H^{s.o.}_{q\bar{q}} 
\end{equation}
where
\begin{equation}
H^{conf}_{q\bar{q}} = C + br + {{\alpha_s(r)} \over r}
\vec{F}_q \cdot \vec{F}_{\bar{q}}  
\end{equation}
includes the spin-independent linear confinement and Coulomb-like
interaction,
\begin{equation}
H^{cont}_{q\bar{q}} = - {{ 8\pi} \over 3} 
{{ \alpha_s(r)} \over{m_{q}m_{\bar{q}} } } \vec{S}_q \cdot \vec{S}_{\bar{q}}
\delta^3 (\vec{r}) \; \vec{F}_q \cdot \vec{F}_{\bar{q}} 
\end{equation}
is the colour contact interaction,
\begin{equation}
H^{ten}_{q\bar{q}} = -{ {\alpha_s(r)}\over {m_q m_{\bar{q}} } } {1\over{r^3}}
\left[{ 
{ { 3\vec{S}_q\cdot \vec{r} \; \vec{S}_{\bar{q}} \cdot \vec{r} } \over {r^2} } 
- \vec{S}_q \cdot \vec{S}_{\bar{q}} 
}\right] \; \; 
\vec{F}_q \cdot \vec{F}_{\bar{q}}
\end{equation}
is the colour tensor interaction,
\begin{equation}
H^{s.o.}_{q\bar{q}} = H^{s.o.(cm)}_{q\bar{q}} + H^{s.o.(tp)}_{q\bar{q}}
\end{equation}
is the spin-orbit interaction with
\begin{equation}
H^{s.o.(cm)}_{q\bar{q}} =-{{\alpha_s(r)}\over {r^3}}
\left({ 
{ {\vec{S}_q} \over {m_q m_{\bar{q}} }} 
+ { {\vec{S}_{\bar{q}} }\over {m_q m_{\bar{q}}} } 
+ { {\vec{S}_q} \over {m_q^2 } } 
+ { {\vec{S}_{\bar{q}} }\over {m_{\bar{q}}^2 } } 
}\right) 
\cdot \vec {L} \; \vec{F}_q \cdot \vec{F}_{\bar{q}} 
\end{equation}
its colour magnetic piece arising from one-gluon exchange and
\begin{equation}
H^{s.o.(tp)}_{q\bar{q}} =- {1\over{2r}}
{ { \partial H_{q\bar{q}}^{conf} } \over{\partial r} }
\left({ { {\vec{S}_q} \over {m_q^2 } } 
+ { {\vec{S}_{\bar{q}} }\over {m_{\bar{q}}^2 } } }\right) \cdot \vec {L}
\end{equation}
the Thomas precession term.  In these formulae, 
$\langle \vec{F}_q \cdot \vec{F}_{\bar{q}} \rangle = -4/3$
for a meson and $\alpha_s(r)$ is the running coupling constant of QCD.  

For the case of a quark and antiquark of unequal mass
the $^3P_1$ and $^1P_1$ states can mix via
the spin orbit interaction or some other mechanism.  
Consequently, the physical $j=1$ states are linear
combinations of $^3P_1$ and $^1P_1$ which we describe by the following mixing:
\begin{eqnarray}
K_{1(low)}^+ & = & (^1P_1)^+ \cos\theta + (^3P_1)^+ \sin\theta 
		= K_b^+ \cos\theta + K_a^+ \sin\theta  \nonumber \\
K_{1(high)}^+ & = & -(^1P_1)^+ \sin\theta + (^3P_1)^+ \cos \theta 
		= -K_b^+ \sin\theta + K_a^+ \cos\theta 
\end{eqnarray}

The  Hamiltonian problem was solved 
using the following parameters: the slope of the
linear confining potential is 0.18 GeV$^2$, 
$m_u=m_d=0.22$ GeV and $m_s=0.419$ GeV.
The resulting masses of the unmixed states are:
\begin{eqnarray}
M(K_a) & = & 1.37 \; \mbox{GeV} \nonumber\\
M(K_b) & = & 1.35 \; \mbox{GeV}.
\end{eqnarray}
We expect these values to be reasonable estimates as 
the model's predictions for the closely related 
$a_1$ and $b_1$ masses are 
consistent with the experimental measurements.
In this model 
spin-orbit mixing results in $\theta_K=-5^o$ \cite{godfrey91} but the
$K_1$ masses remain the same 
within the given numerical precision.  
These mixed masses and the 
mixing angle are not consistent with the measured values.

We can obtain a phenomenological estimate of $\theta_K$ by 
considering the $2 \times 2$ matrix relating $K_a$ and $K_b$ to the 
physical $K_1$'s.  We do not make any assumptions about the origin of 
the $^3P_1 - ^1P_1$ mixing and treat the off-diagonal matrix element of the 
$K_1$ mass matrix as a free parameter.
Diagonalizing the $K_a-K_b$ mass matrix gives the relation 
between $\theta_K$ and the mass differences:
\begin{equation}
\cos 2\theta_K = 
{ { M(K_a)-M(K_b) } \over  { M (K_1(1402)) - M(K_1(1273)) } }
\end{equation}
with corresponding $K_1$ masses:
\begin{eqnarray}
M_{low} & = & M_b \cos^2\theta_K + M_a \sin^2\theta_K - (M_a - M_b) 
{{\sin^2 2\theta_K}\over{2 \cos 2\theta_K}} \nonumber \\
M_{high} & = & M_b \sin^2\theta_K + M_a \cos^2\theta_K + (M_a - M_b) 
{{\sin^2 2\theta_K}\over{2 \cos 2\theta_K}}. 
\end{eqnarray}
Solving gives $\theta_K \simeq \pm 41^o $.  Note that degenerate $K_a$ 
and $K_b$ masses will always result in a mixing angle of $\pm 45^o$
\cite{mixing}.  
Thus, the value we obtain for $\theta_K$ is more a reflection of the 
near degeneracy of the model's prediction for $M_{K_a}$ and $M_{K_b}$ 
than anything else and one should not read too much into 
the value we extract here.

\section{WEAK COUPLINGS OF THE $K_1$'s}

We use the mock meson approach to calculate the hadronic matrix elements
\cite{godfrey85,capstick90,yaouanc74,hayne82,godfrey86,isgw89}.  
The basic assumption of the mock meson approach is that physical hadronic
amplitudes can be identified with the corresponding quark model amplitudes
in the weak binding limit of the valence quark approximation.  
This correspondence is exact only in the limit of zero binding and in
the hadron rest frame.  Away from this limit the amplitudes are not in
general Lorentz invariant by terms of order $p_i^2 /m_i^2$.
In this approach the mock meson, which we denote by $\widetilde{M}$, is 
defined as a state of a free quark and antiquark with the wave function of the
physical meson, $M$:
\begin{equation}
\vert \widetilde{M}(\vec K ) \rangle = \sqrt{2M_{\widetilde{M}} }
\int d^3p \; \Phi_M (\vec{p} ) \chi_{s\bar s} \phi_{q\bar q} 
\phi_{colour}
\vert q[ (m_q/\mu )\vec{K} + \vec{p},s]\; \bar{q} [(m_{\bar q}/\mu) \vec{K}
-\vec{p}, \bar{s} ]\rangle 
\end{equation}
where $\Phi_M(\vec{p})$, $\chi_{s\bar{s}}$, $ \phi_{q\bar q}$, and
$\phi_{colour}$ 
are momentum,
spin, flavour, and colour
wave functions respectively, $\mu=m_q +m_{\bar q}$, 
$\vec{K}$ is the mock meson momentum, 
$M_{\widetilde{M}}$ is the mock meson mass,
and $\sqrt{2M_{\widetilde{M}}}$ is included to normalize the mock 
meson wavefunction.
To calculate the hadronic amplitude, the physical
matrix element is expressed in terms of Lorentz covariants with 
Lorentz scalar coefficients $A$.  In the simple cases when the mock-meson 
matrix element has the same form as the physical meson amplitude
we simply take $A$ = $\tilde A$.  

In the case of interest, 
the axial meson decay constants are expressed as:
\begin{equation}
\langle 0 \vert \bar{q} \gamma^{\mu} (1-\gamma_5 ) q \; \vert
\; M(\vec{K},\lambda) \rangle = {i\over {(2\pi )^{3/2} } } f_{K_1} 
\epsilon^{\mu}(\vec{K},\lambda)
\end{equation}
where $\epsilon^{\mu}(\vec{K},\lambda)$ is the $K_1$ polarization 
vector and $f_{K_1}$ is the appropriate $K_1$ decay constant.
To calculate the left hand side of Eq. (13) we first calculate
\begin{equation}
\langle 0 \vert \bar{q} \gamma^{\mu} (1-\gamma_5 ) q
\;\vert q[(m_q /\mu) \vec{K} + \vec{p},s]\; \bar{q} [(m_{\bar q}/\mu) \vec{K} 
-\vec{p}, \bar{s} ]\rangle 
\end{equation}
using free quark and antiquark wavefunctions and
weight the result  with the meson's momentum space wavefunction.

There are a number of ambiguities in the mock-meson approach and 
different prescriptions have appeared in the literature.
For example, there are several different definitions for the mock-meson mass 
($M_{\widetilde{M}}$) appearing in Eq. 12.
To be consistent with the mock meson prescription, 
we should use the mock meson
mass defined as $\langle E_q \rangle + \langle E_{\bar{q}} \rangle$. 
However, because it is introduced  to
give the correct relativistic normalization of the meson's wavefunction
the physical mass is another, perhaps more appropriate, definition.
The second ambiguity is the question of which component of the 4-vector
in Eq. 13 we should use to obtain $f_{K_1}$.
In principle, it should not matter as both the left and right sides of 
Eq. 13 are Lorentz 4-vectors.  
This is true in  the weak binding limit where binding effects are
totally neglected, but in 
practice, this is not the case.  We follow Ref. \cite{capstick90}
and extract $f_{K_1}$ using the spatial components of Eq. 13 in the limit
$\vec{K}\to 0$.  Finally, evaluating Eq. 14 introduces factors of 
$\sim m_i/E_i$.  While some prescriptions take the expression 
derived from Eq. 13 only as a guideline and introduce powers of 
$\sim (m_i/E_i)^\epsilon$ with $\epsilon$ an arbitrary power, we 
chose to use the expression exactly as derived from take Eq. 13.  
The different prescriptions are described in greater detail 
in Ref. \cite{capstick90} which calculated the 
pseudoscalar decay constants ($f_K$).  We will 
follow the approach taken there and use the variations in 
prescriptions as a measure of how seriously we should take our 
results.  In our results we therefore use the ``exact'' expression for 
$f_{K_1}$ and we take $M_{\widetilde{M}}$ to be equal to the physical mass
($M_{phys}$).
Variations in the mock-meson normalization
result in variations in $f_{K_1}$ of at most 20\%.
Results using the physical mass lie in the middle of the range so 
that we expect uncertainties introduced by taking $M_{\widetilde{M}}\equiv 
M_{phys}$ to be no more than $\sim 10\%$. 
As in Ref. \cite{capstick90} $f_{K_1}$ was most sensitive to the 
wavefunction used.  Here we use the sets of wavefunctions that gave 
the best agreement with experiment for $f_K$ in Ref. \cite{capstick90}. 
We choose two possibilities, one 
which underestimated $f_K$ and one which 
overestimated it.  We would expect these choices to likewise bound 
the actual value of the $f_{K_1}$.

The expressions we obtain for $f_{K_1}$ are given by:
\begin{eqnarray}
f_{K_1} (^3P_1)  & = & -{{ 4 \sqrt{3}}\over 3} 
\sqrt{M_{\widetilde{K}_1}}  
\int {{ d^3p} \over {(2\pi)^{3/2}} }
\left( i \sqrt{3\over 8 \pi} \phi_{K_1} (p) \right) \; p \; \nonumber \\
& & \qquad \quad \times \left[ \left( { {E_q+m_q} \over {2E_q} } \right) 
\left( { {E_{\bar{q}} +m_{\bar{q}} }\over { 2 E_{\bar{q}} }  } 
\right) \right]^{1/2}
\left[ { 1 \over {E_q +m_q } } + { 1\over {E_{\bar{q}} +m_{\bar{q}} } }
\right] 
\nonumber\\
f_{K_1} (^1P_1)  & = & +{{ 2 \sqrt{6}}\over 3} 
\sqrt{M_{\widetilde{K}_1}}  
\int {{ d^3p} \over {(2\pi)^{3/2}} }
\left( i\sqrt{3\over 8 \pi} \phi_{K_1} (p) \right) \; p \; \nonumber \\
& & \qquad \quad \times \left[ \left( { {E_q+m_q} \over {2E_q} } \right) 
\left( { {E_{\bar{q}} +m_{\bar{q}} }\over { 2 E_{\bar{q}} }  } 
\right) \right]^{1/2}
\left[ { 1 \over {E_q +m_q } } - { 1\over {E_{\bar{q}} +m_{\bar{q}} } }
\right] 
\end{eqnarray}
where $\phi_{K_1}(p)$ is the radial part of the momentum space
wavefunction, $E_q = \sqrt{|\vec{p}|^2 +m_q^2 }$ and
$E_{\bar{q}} = \sqrt{|\vec{p}|^2 +m_{\bar{q}}^2 }$. 
In the $SU(3)$ limit only $K_a$ couples to the weak current.

With the definition of $f_{K_1}$ given by Eq. (13) the partial 
width for $\tau \to K_1 \nu_\tau$ is given by:
\begin{equation}
\Gamma (\tau \to K_1 \nu_\tau ) 
= {{G_F^2 |V_{us}|^2 f_{K_1}^2 m_\tau^3}\over {16\pi m_{K_1}^2}}
(1-m_{K_1}^2/m_\tau^2)^2 (1 + 2m_{K_1}^2/m_\tau^2)
\end{equation}

\subsection{The Non-relativistic Limit}

It is useful to examine the $K_1$ decay constants in the 
non-relativistic limit where their qualitative properties are more 
transparent.  In this limit the axial-vector meson decay constants become:
\begin{eqnarray}
f_{K_1} (^3P_1)  & = & - \sqrt{12 M_{\widetilde{K_1}} } 
\left[ { {1\over m_q} + {1\over m_{\bar{q}}}  } \right] 
\left. \sqrt{3 \over{8\pi}}  {{\partial R_P (r)}\over {\partial r}} 
\right|_{r=0}  \nonumber \\
f_{K_1} (^1P_1) & = & \sqrt{6M_{\widetilde{K_1}} }
\left[ { {1\over m_q} - {1\over m_{\bar{q}}}  } \right]
\left. \sqrt{3 \over{8\pi}}  {{\partial R_P (r)}\over {\partial r}} 
\right|_{r=0}
\end{eqnarray}
where 
$R_P(r)$ is the radial part of the coordinate space wavefunction.
Combining the weak decay amplitudes with the mixed $K_1$ eigenstates 
the decay constants for the mixed $|K^+ \rangle = 
-|u\bar{s}\rangle$ states are given by:\footnote{Note that the sign 
change going from Eqn. 17 to Eqn. 18 comes from the phase in the 
$K^+$ flavour wavefunction.}
\begin{eqnarray}
f_{K_{low}}  & = & - A
\left[\left( { {1\over m_u} - {1\over m_s}  } \right) \cos\theta_K
-\sqrt{2}  \left( { {1\over m_u} + {1\over m_s}  } \right) \sin\theta_K
\right]
  \nonumber \\
f_{K_{high}}  & = & + A
\left[\left( { {1\over m_u} - {1\over m_s}  } \right) \sin\theta_K
+\sqrt{2}  \left( { {1\over m_u} + {1\over m_s}  } \right) \cos\theta_K
\right]
\end{eqnarray}
where we have defined 
\begin{equation}
A=\sqrt{6 M_{\widetilde{K_1}} } 
\left( \left. \sqrt{3 \over{8\pi}}  {{\partial R_P (r)}\over {\partial r}} 
\right|_{r=0} \right).
\end{equation}

In the SU(3) limit $f_{K_1}(^1P_1)$ explicitly goes 
to zero and only the $^3P_1$ state couples to the weak current.
The $K_b$ coupling therefore goes like the $SU(3)$ breaking $(m_s - m_d)$.

\subsection{Extracting $\theta_K$ Using the Non-Relativistic 
Expressions}

We can obtain an estimate of the $^3P_1 -^1P_1$ mixing angle by 
comparing the quark model predictions to experiment.  As stated 
above, the values of the decay constants were quite sensitive to the 
choice of wavefunction.  We calculated the $f_{K_1}$ for two sets of 
wavefunctions that gave the best agreement between a quark model 
calculation and experiment for the pseudoscalar decay 
constants\cite{capstick90}. We expect that the actual values for the 
$f_{K_1}$ will lie between the values predicted using these 
wavefunctions. The values for the two meson masses and 
two sets of wavefunctions are given in Table I.

There are four measurements that can be used to constrain $\theta_K$.
The TPC/Two-Gamma Collaboration \cite{tpc95} has made the measurements:
\begin{eqnarray}
BR (\tau \to \nu K_1(1270)) & = & (0.41^{+.41}_{-.35} )\times 10^{-2} \\
BR (\tau \to \nu K_1(1400)) & = & (0.76^{+.40}_{-.33} )\times 10^{-2} \\
BR (\tau \to \nu K_1) & = & (1.17^{+.41}_{-.37} )\times 10^{-2} 
\end{eqnarray}
and Alemany \cite{alemany94} combines CLEO  and ALEPH data 
\cite{cleo} to 
obtain:
\begin{equation}
BR (\tau \to \nu K_1)  = (0.77 \pm 0.12)  \times 10^{-2} 
\end{equation}
which is smaller than, but consistent with, the TPC/Two-Gamma
result.  CLEO claims that the $\tau$ decays preferentially to the 
$K_1(1270)$.   

Using the ratio $BR (\tau \to \nu K_1(1270))/BR (\tau \to \nu 
K_1(1400))$ has the advantage of factoring out the uncertainties 
associated with the $K_1$ wavefunction.  The ratio is given 
by\footnote{The numbers from Table I give a 
slightly different value since the different $K_1$ masses in our 
expression for $f_{K_1}$ do not exactly factor out.}
\begin{equation}
R= 1.83 \left| {{\sin\theta_K - \delta \cos\theta_K}
\over {\cos\theta_K + \delta \sin\theta_K}} \right|^2
\end{equation}
where 1.83 is a phase space factor and 
$\delta$ is an $SU(3)$ breaking factor given by
\begin{equation}
\delta = {1\over \sqrt{2}} \left( {{m_s-m_u}\over {m_s + m_u}} \right).
\end{equation}
The ratio, R, is plotted in Fig.~1a as a function of $\theta_K$.
Taking $m_u =0.33$~GeV and $m_s=0.55$~GeV and fitting Eq.~(24) to the 
ratio of the TPC/Two-Gamma results we obtain $-30^o \lesssim \theta_K \lesssim 
50^o$ at 68 \% C.L. 
where the large uncertainty is directly attributed to the large 
errors in the branching ratios.\footnote{ The $\chi^2$ of the fit 
actually has 2 local minima corresponding to both a negative and 
positive solution.  However, since the hump separating the two 
solutions is approximately equal to $\Delta \chi^2 \leq 1$ the 
entire range given for $\theta_K$ is consistent at 68 \% C.L.}

Although the relative errors for the individual branching ratios are smaller 
than those of the ratio, especially for the sum to the two $K_1$ 
states, using the branching ratios introduces additional
uncertainties due to the errors associated with the meson 
wavefunction.  In addition, the branching ratios turn out to be 
less sensitive to $\theta_K$ than the ratio.  This is seen very 
clearly in Figs.~1b, 1c, and 1d where we have plotted the branching 
ratios for $\tau \to \nu_\tau K_1 (1270)$, $\tau \to \nu_\tau 
K_1(1400)$ and the sum of the two respectively. The values 
$\tau_\tau=(295.6 \pm 3.1)\times 10^{-15}$~s and $|V_{us}|=0.2205 
\pm 0.0018$ were used to obtain these curves \cite{pdb}.
The two curves in 
each figure represent the two wavefunctions we use and we have 
included the experimental value with its error.  In Fig.~1d both 
the TPC/Two-Gamma and the CLEO/ALEPH values are shown.
It is apparent from these figures that it is not particularly 
meaningful to extract a value for $\theta_K$ from these results and 
any value would be very model dependent.  
Clearly better data is needed.  The ratio of the 
rates into the individual final states 
will give the most model independent constraints on $\theta_K$.

\subsection{Extracting $\theta_K$ Using the Relativized Expressions}

We next calculate the axial meson decay constants using the
relativized formula of Eq.~(15).   
One might question the importance of including
relativistic corrections.  However, we need only consider the importance of
another relativistic correction:  QCD hyperfine interactions which
give rise to the $\rho-\pi$, $K^* -K$, $\ldots$, $B^*-B$ splittings 
\cite{rqm}.  
Although it is difficult to gauge the importance of relativistic 
corrections to the $f_{K_1}$, if nothing else their inclusion 
acts as one more means of judging the reliability of the results.

As in the previous section we give results for two wavefunction sets 
that give reasonable agreement for the $f_K$ in a similar 
calculation.  The various $f_{K_1}$ are given in Table II.  
We expect that the 
actual values will lie between the two values given for each 
case.  The predictions for the various branching fractions are shown 
in Fig.~2 as a function of $\theta_K$ along with the experimental 
values.  The most reliable constraint again comes from the ratio of 
branching fractions which gives $-35^o \lesssim \theta_K \lesssim 45^o$ at 
68 \% C.L.
One could also extract values using $BR(\tau\to \nu K_1(1400))$ 
and $BR(\tau\to \nu K_1(1270))$ but as in the non-relativistic case 
these values are quite sensitive to the 
magnitude of $f_{K_1}$ which depends on the poorly known $K_1$ 
wavefunctions.  

We conclude that the decays $\tau \to \nu K_1$ offer a means of 
measuring the $^3P_1-^1P_1$ mixing angle but to do so will require 
more precise measurements than are currently available.

\section{STRONG DECAYS OF THE $K_1$'s}

It is well known that the strong decays of the $K_1$ mesons provides 
a means of extracting the $^3P_1 -^1P_1$ mixing angle\cite{carnegie77}.  
In particular the $BR(K_1(1270)\to K^* \pi)/BR(K_1(1400)\to K^* \pi)$ and 
$BR(K_1(1270)\to K \rho)/BR(K_1(1400)\to K \rho)$ ratios have been 
especially useful.  We examine the decays to the final states $K \rho$, 
$K \omega$, and $K^* \pi$.  Although other decays are observed they lie 
below threshold and proceed through the tails of the Breit-Wigner 
resonances making the calculations less reliable.
In this section we examine the strong $K_1$ 
decays using the pseudoscalar emission model \cite{godfrey85}, 
the $^3P_0$ model (also known as the 
quark-pair creation model) \cite{yaouanc73},
and the flux-tube breaking model \cite{kokoski}.  
We concentrate on the decays $K_1\to K^*\pi$ and $K_1 \to \rho K$.  

For the decays $K_1 \to VP$, where $V$ and $P$ denote vector and 
pseudoscalar mesons respectively, the OZI-rule-allowed decays can be 
described by two independent $S$ and $D$-wave amplitudes which we 
label $S$ and $D$.  The decay amplitudes, using the conventions of 
Eq.~(8) are given by
\begin{eqnarray}
A (K_1^{low} \to [K^* \pi]_S) 
& = &  -S \sin(\theta_K - \theta_0 ) \nonumber \\
A (K_1^{low} \to [K^* \pi]_D) 
& = & +D \cos(\theta_K - \theta_0 ) \nonumber \\
A (K_1^{high} \to [K^* \pi]_S) 
& = & -S \cos(\theta_K - \theta_0 ) \nonumber \\
A (K_1^{high} \to [K^* \pi]_D) 
& = & -D \sin(\theta_K - \theta_0 ) \nonumber \\
A (K_1^{low} \to [\rho K]_S) 
& = & +S \sin(\theta_K + \theta_0 ) \nonumber \\
A (K_1^{low} \to [\rho K]_D) 
& = & +D \cos(\theta_K + \theta_0 ) \nonumber \\
A (K_1^{high} \to [\rho K]_S) 
& = & +S \cos(\theta_K + \theta_0 ) \nonumber \\
A (K_1^{high} \to [\rho K]_D) 
& = & -D \sin(\theta_K + \theta_0 ) \nonumber\\
A (K_1^{high} \to [\omega K]_S) 
& = & +\sqrt{\frac{1}{3}} S \cos(\theta_K + \theta_0 ) \nonumber \\
A (K_1^{high} \to [\omega K]_D) 
& = & -\sqrt{\frac{1}{3}} D \sin(\theta_K + \theta_0 ) \label{sddef}
\end{eqnarray}
where $\sin\theta_0=\sqrt{1/3}$ and $\cos\theta_0=\sqrt{2/3}$ and 
the subscripts $S$ and $D$ refer to $S$- and $D$-wave decays.
In the heavy quark limit the $j=1/2$ state decays into $K^* \pi$ in 
an S-wave and the $j=3/2$ state decays into $K^* \pi$ 
in a D-wave.  Since the decay $K_1(1400)\to 
K^*\pi$ is dominantly $S$-wave while the decay $K_1(1270)\to K^*\pi$ 
has comparable $S$ and $D$-wave contributions we conclude that
experimental data favours the 
heavier $J^P = 1^+$ to be mainly $j=1/2$ and the lighter one to be 
mainly $j=3/2$.

In the following sections we give results for these amplitudes, the 
resulting decay widths and the fitted values of $\theta_K$ for the 
various decay models.  

\subsection{Decays by the Pseudoscalar-Meson Emission Model}

In this approach meson decay proceeds through a single-quark 
transition via the emission of a pseudoscalar meson \cite{godfrey85}.
We assume that the pair creation of $u$, $d$, and $s$ quarks is 
approximately $SU(3)$ symmetric.  We follow Ref. \cite{godfrey85} 
and use the various approximations introduced there.  The resulting 
amplitudes are given by
\begin{eqnarray}
D & = & {1\over 2} A \tilde{q}^2 F(q^2) \\
S & = & \tilde{S} F(q^2)
\end{eqnarray}
where $A=1.67$, $\tilde{S}=3.27$, $q$ is the momentum of each
outgoing meson in the centre of mass (CM) frame, 
$\tilde{q}=q/\beta$, $\beta=0.4$~GeV and
\begin{equation}
F(q^2) = \sqrt{1 \over 2}
 \left( {q\over 2\pi}\right)^{1/2} \exp (-q^2/16\beta^2).
\end{equation}
Numerical values for the amplitudes are given in Table III.

The partial widths for $K_1\to K^*\pi$ and $K_1 \to \rho K$ and the 
ratio of the $D$ to $S$ amplitudes for $K_1 \to K^* \pi$ are plotted 
in Fig.~3 as a function of $\theta_K$ for the $K_1(1270)$ and 
$K_1(1400)$.  The experimental values are given with their errors. 
From the figures it is clear that the experimental values for
$K_1(1270)\to K^*\pi$, $K_1 (1400) \to \rho K$, and
$A_D/A_S[K_1(1400)]$ correspond to 
minima in the quark model results with $\theta_K \sim 45^o$.  We 
performed a $\chi^2$ fit to the data listed in Table IV
and obtained $\theta_K=48^o\pm 5^o$.  We 
also allowed the $\tilde{S}$, $A$, and $\beta$ parameters to vary  and obtained 
very similar results,  the main difference being that the $\chi^2$ 
value at the minimum decreased significantly.  The partial widths 
and $A_D/A_S (K_1\to K^*\pi)$ 
ratios are given in Table IV for the fitted value of 
$\theta_K$.

\subsection{Decays by the Flux-Tube Breaking Model}

The flux-tube breaking model is a variation of the 
$^3P_0$ model which more closely describes the actual 
decay processes.  In the $^3P_0$ model the elementary 
process is described by the creation of a $q\bar{q}$ pair with the 
quantum numbers of the vacuum, $J^{PC}=0^{++}$.  
The greatest advantage of this approach is that it requires only one 
overall normalization constant for the pair creation process.
In the flux-tube breaking model, the flux-tube-like structure of the 
decaying meson and its implications for $^3P_0$
amplitudes are taken into account by viewing a meson decay as 
occurring via the breaking of the flux-tube with the simultaneous 
creation of a quark-antiquark pair.  To incorporate this into the 
$^3P_0$ model, the pair creation amplitude $\gamma$ is allowed to vary 
in space so that the $q\bar{q}$ pair is produced within the confines 
of a flux-tube-like region surrounding the initial quark and 
antiquark.  This model is described in detail in Ref. \cite{kokoski}.
The $^3P_0$ model corresponds to the limit in which $\gamma$ is 
constant.

For the $^3P_0$ model using simple harmonic oscillator wavefunctions
the $S$ and $D$ amplitudes are given by:
\begin{eqnarray}
S & = & \left[ 3 - 
q^2 \left( {{m_{13}\beta_B^2 + m_{23}\beta_C^2}\over {3\beta_B^4 
\beta_C^4}} \right) 
\left[ 3 \beta_B^2 \beta_C^2 - \beta^2 (m_{13}\beta_B^2 + 
m_{23}\beta_C^2) \right] \right] F(q^2) A \\
D & = &  
q^2 \left( {{m_{13}\beta_B^2 + m_{23}\beta_C^2}\over {3\beta_B^4 
\beta_C^4} } \right) 
\left[ 3 \beta_B^2 \beta_C^2 - \beta^2 
(m_{13}\beta_B^2 + m_{23}\beta_C^2) \right] F(q^2) A
\end{eqnarray}
where
\begin{eqnarray}
F(q^2) & = & \exp \left[ {-q^2 \over 6} 
\left( {{\beta^2 [(m_{13}-m_{23})^2 \beta_A^2 + m_{13}^2 \beta_B^2
+m_{23}^2 \beta_C^2 ]}\over{\beta_A^2 \beta_B^2 
\beta_C^2}}\right)\right] \\
A & = & {{2i\gamma} \over {27 \pi^{1/4}\beta^{1/2} } }
\left({\beta\over \beta_A}\right)^{5/2}
\left({\beta^2 \over \beta_B \beta_C}\right)^{3/2} q^{1/2} 
\left( { {\widetilde{M}_B \widetilde{M}_C } 
\over {\widetilde{M}_A} } \right) ^{1/2} \\
\beta^{-2} & = & {1\over 3} (\beta_A^{-2} +\beta_B^{-2} +\beta_C^{-2}) \\
m_{13} &=& {m_1\over m_1 + m_3} \\
m_{23} &=& {m_2\over m_2 + m_3}, 
\end{eqnarray}
$m_1$ and $m_2$ are the quark and antiquark masses from the original 
meson, $m_3$ is the mass of the created quark/antiquark,  the $\beta_i$ are 
the simple harmonic oscillator wavefunction parameters, and $q$ is 
the momentum of each outgoing meson in the CM frame.
For these results we take the $\widetilde{M}_i$ to be equal to
the calculated masses of the mesons in a 
spin-independent potential \cite{kokoski}.
Numerical values for the relevant amplitudes are given in Table III.

The decay amplitudes in the $^3P_0$ model were computed symbolically using
Mathematica \cite{wolfram}.  In the flux-tube breaking model two of the six
integrals were done analytically; the remaining four were done numerically.
The integrands were prepared symbolically using Mathematica and then integrated
numerically using either adaptive Monte Carlo (VEGAS \cite{vegas}) or a
combination of adaptive Gaussian quadrature routines.

We calculated the $K_1$ strong decays using both the flux-tube 
breaking model and the $^3P_0$ model for several sets of 
wavefunctions.  In all cases we fitted $\gamma$ to 28 of the best 
known meson decays by minimizing the $\chi^2$ defined by $\chi^2 
=\sum_i (\Gamma^{theory}_i - \Gamma^{exp}_i)^2/\delta\Gamma_i^2$ where 
$\delta\Gamma_i$ is the experimental error \footnote{For the calculations in
the flux-tube breaking model, a 1\% error due to the numerical integration was 
added in quadrature with the experimental error.}. The details of these 
fits are given in Ref. \cite{blundell95}. 
We performed a second fit to
the $K_1$ decays where we allowed both $\theta_K$ 
and $\gamma$ to vary.  
The value of $\theta_K$ obtained in the 
second approach did not change much from the first value --- the main 
difference was that the $\chi^2$ in the second fit was reduced 
substantially.  The values for $\gamma$ obtained in the second set 
of fits are consistent, 
within errors, with those obtained by the global fit of Ref. 
\cite{blundell95}.
In Fig.~4  we show the decay widths 
and ratios of D to S amplitudes as a function of 
$\theta_K$ for the $^3P_0$ model.  
The results for the two variations of the flux-tube breaking model are very 
similar and are therefore not shown.
It is clear from these figures that 
$\theta_K$ will be approximately equal to $45^o$.  
The fitted values of $\theta_K$ for the various models, and the resulting 
widths, are given in Table IV.

\section{DISCUSSION}

One of the motivations for this analysis is to relate hadron 
properties to the underlying theory via {\it effective} 
interquark interactions \cite{schnitzer89}.  
We begin our discussion of the $K_1$ mesons by  rewriting the
non-relativistic spin dependent potential in a more suitable form
and interpreting it as an {\it effective} interaction \cite{schnitzer89}.  
We will later examine the $K_1$ meson properties in the limit $m_Q \to 
\infty$.

The spin-orbit Hamiltonian can be rewritten as:
\begin{eqnarray}
H_{q\bar{q}}^{s.o.(cm)} & = & {2\over 3} {{\alpha_s}\over{r^3}}
\left( {1\over m_q} + {1\over m_{\bar{q}}} \right)^2
 \vec{S} \cdot \vec{L} 
+ {2\over 3} {{\alpha_s}\over{r^3}} 
\left( {1\over m_q^2} - {1\over m_{\bar{q}}^2 } \right)
\vec{S}_-
\cdot \vec{L} \\
H_{q\bar{q}}^{s.o.(tp)} & = &
- {1\over 4r } { { \partial H_{q\bar{q}}^{conf} } \over{\partial r} }
\left[ \left( {1\over m_q^2} + {1\over m_{\bar{q}}^2} \right)
\vec{S} \cdot \vec{L} 
+ \left( {1\over {m_q^2}} - {1\over m_{\bar{q}}^2} \right) 
\vec{S}_-
\cdot \vec{L}  \right]  
\end{eqnarray}
where  $\vec{S}= \vec{S}_q +\vec{S}_{\bar{q}}$, 
$\vec{S}_- = \vec{S}_q -\vec{S}_{\bar{q}}$.
Taking $\bar{q}=Q$ the various terms in $H^{s.o.}$ can be rearranged as
\begin{eqnarray}
H^{s.o.} & = & H_{s.o.}^+ \; \vec{S}\cdot \vec{L} + H_{s.o.}^- 
\; \vec{S}_-\cdot \vec{L} \nonumber \\
& = & H_{s.o.}^q \; \vec{S}_q\cdot \vec{L} + H_{s.o.}^Q 
\; \vec{S}_Q\cdot \vec{L}
\end{eqnarray}
where the definitions of $H_{s.o.}^+$, $H_{s.o.}^-$, $H^q_{s.o.}$, and 
$H_{s.o}^Q$ follow from Eqs.~(37) and (38).  
It is the $H^-_{s.o}$ term which gives rise to the spin-orbit mixing 
between the singlet and triplet states.
With this Hamiltonian, we obtain the following mass formulae 
for the P-wave mesons:
\begin{eqnarray}
M(^3P_2) & = & M_0 +{1\over 4} \langle H_{cont} \rangle - {1\over{10}}
\langle H_{ten} \rangle + \langle H^+_{s.o.} \rangle \nonumber\\
\pmatrix{ M(^3P_1) \cr M(^1P_1) \cr} 
& = & \pmatrix{ M_0 +{1\over 4} \langle H_{cont} \rangle + {1\over 2}
\langle H_{ten} \rangle - \langle H^+_{s.o.} \rangle 
& \sqrt{2} \langle H^-_{s.o.} \rangle \cr
\sqrt{2} \langle H^-_{s.o.} \rangle   &
M_0 -{3\over 4} \langle H_{cont} \rangle \cr}
\; \pmatrix{ ^3P_1 \cr ^1P_1 \cr} \nonumber\\
M(^3P_0) & = & M_0 +{1\over 4} \langle H_{cont} \rangle - 
\langle H_{ten} \rangle -2  \langle H^+_{s.o.} \rangle 
\end{eqnarray}
where the $\langle H_i \rangle$
are the expectation values of the spatial parts of the various
terms, $M_0$ 
is the center of mass of the multiplet, and we have adopted a phase
convention corresponding to the order of coupling 
$\vec{L} \times \vec{S}_q \times \vec{S}_Q $.

We can rewrite $H^{s.o.}$ using the substitutions $\bar{m}={1\over 
2} (m_q + m_Q)$ and $\Delta = (m_Q - m_q )$ to obtain the 
approximate expression
\begin{equation}
H^{s.o.}  \simeq  \left[ {8\over 3} {\alpha_s \over \bar{m}^2 r^3}
- {1\over 2\bar{m}^2} {1\over r} 
{ { \partial H_{q\bar{q}}^{conf} } \over{\partial r} } \right]
 \vec{S} \cdot \vec{L}  
 - {\Delta \over \bar{m} } 
\left[ {4\over 3} {\alpha_s \over \bar{m}^2 r^3}
- {1\over 2\bar{m}^2} {1\over r} 
{ { \partial H_{q\bar{q}}^{conf} } \over{\partial r} } \right]
 \vec{S}_- \cdot \vec{L} .
\end{equation}
Written in this way one sees that there is a factor of two 
difference between the colour magnetic term and the Thomas 
precession term for the $H_{s.o}^-$ relative to $H_{s.o}^+$.  
The observed spin-orbit splittings in hadrons indicate a delicate 
cancellation between the colour magnetic and Thomas precession 
spin-orbit terms.  Given this cancellation, the factor 
of two could lead to a large effect or even a sign reversal in the 
spin-orbit mixing.

In particular, the relativized quark model gives $\theta_K = -5^o$ 
\cite{godfrey91}.  This originates from $\langle H_{s.o.}^- \rangle 
\sim -1$~MeV.  On the other hand, the various phenomenological 
measurements give $\theta_K \sim 40^o$ which implies
a value of $\langle H_{s.o.}^- \rangle \sim 40$~MeV.  Comparing 
these numbers to $\langle H_{s.o.}^+ \rangle \sim 47$~MeV 
extracted from Ref. \cite{godfrey91} one can 
see that by extracting a value for $\langle H_{s.o.}^- \rangle$ from 
$\theta_K$ and comparing it to the value for 
$\langle H_{s.o.}^+ \rangle$ one can obtain 
information about the relative strengths of the Coulomb and 
confining pieces of $H_{q\bar{q}}^{conf}$.
Given the sensitivity of the mixing angle to the delicate 
cancellation between terms, $L-S$ mixing can therefore be a useful means of 
probing the confinement potential.\footnote{Nevertheless, 
one cannot rule out the possibility that another mechanism is 
responsible for $^3P_1$-$^1P_1$ mixing such as mixing via common decay 
channels \cite{lipkin77}.}

We next consider the heavy quark limit,
where $m_Q\to \infty$.  In this limit the mass formulae simplify to:
\begin{eqnarray}
M(^3P_2) & = & M_0 + \langle H^q_{s.o.} \rangle \nonumber \\
\pmatrix{ M(^3P_1) \cr M(^1P_1) \cr} 
& = &  \pmatrix{ M_0 - \langle H^q_{s.o.} \rangle 
& \sqrt{2} \langle H^q_{s.o.} \rangle \cr
\sqrt{2} \langle H^q_{s.o.} \rangle   & M_0  \cr}
\; \pmatrix{ ^3P_1 \cr ^1P_1 \cr} \nonumber \\
M(^3P_0) & = &M_0  -2  \langle H^q_{s.o.} \rangle 
\end{eqnarray}
The two mixed $K_1$ mass eigenstates of $J^P=1^+$ 
appropriate to the heavy quark limit are
described by the total angular momenta $j$ of the light quark with 
$j=1/2$ and $j=3/2$ which are degenerate with the $J^P=0^+$ and 
$J^P=2^+$ states respectively. 
In what follows we will take $\langle H^q_{s.o.} \rangle$ positive
but similar results are obtained for $\langle H^q_{s.o.} \rangle $ 
negative.  For $\langle H^q_{s.o.} \rangle>0 $
the mixing angle is given by
$\sin\theta_K = -\sqrt{2/3}$ and $\cos\theta_K=\sqrt{1/3}$
($\theta_K=-54.7^o$)
with $M_{K_{low}}$ degenerate with the $^3P_0$ ($j=1/2$) state
and $M_{K_{high}}$ degenerate with the $^3P_2$ ($j=3/2$) state.  

For the $K_1$ decay constants,
in the limit that $m_{\bar{s}}$ becomes 
infinitely heavy, the $f_{K_1}$ become proportional to the inverse of 
the light quark mass and are given by 
\begin{eqnarray}
f_{K_{low}} & = & {\sqrt{3} A \over m_u} \nonumber \\
f_{K_{high}} & = & 0
\end{eqnarray}
So in the heavy quark limit 
only the $j=1/2$ state couples to the weak current.  
By comparing this result to the measured decays one 
might learn how well the heavy quark limit describes the strange 
axial mesons.
Using the value of $\theta_K$ that gives the $j=1/2$ and $j=3/2$ 
eigenstates (expected in the heavy quark limit)
and using a finite 
mass strange quark (still taking $\langle H^q_{s.o.} \rangle >0$)
the decay constants are given by:
\begin{eqnarray}
f_K (j=1/2) & = & +{\sqrt{3} A\over m_u} (1+{m_u\over 3m_s}) \nonumber \\
f_K(j=3/2) & = & -\sqrt{8\over 3} {A\over m_s}.  
\end{eqnarray}
The value does not change very much for the $j=1/2$ state, 
$m_u/3m_s \simeq 0.2$, but the  $j=3/2$ state decay constant is no 
longer zero but is now similar in magnitude to that of the $j=1/2$ state.

More importantly, the $\theta_K$ we used in the above 
discussion was based on the $J^P=1^+$ mass matrix obtained for the 
heavy quark limit which  assumes that the contact and tensor 
contributions are negligible.  However,  values for these terms 
extracted from predictions of the relativized quark model 
\cite{godfrey91} are: $\langle H_{cont} \rangle =33$~MeV,
$\langle H_{ten} \rangle =56$~MeV, and
$\langle H^+_{s.o.} \rangle =47$~MeV.  
Clearly the assumption that the contact and tensor pieces are 
negligible is not supported by this model  so that the 
heavy quark limit is questionable for the $s$ quark.

We conclude that while the heavy quark limit is an interesting 
means of making qualitative observations the actual situation for 
the strange axial mesons is far more complicated.

\section{CONCLUSIONS}

In this paper we studied the properties of the strange axial mesons 
in the quark model.  We extracted the $K_1 (^3P_1) - K_1 (^1P_1)$ 
mixing using the mass predictions, by comparing a quark model 
calculation of the $K_1$ decay constants to the decays $\tau \to 
\nu_\tau K_1$ and by comparing strong decay widths calculated using 
the pseudoscalar emission model and the flux-tube breaking model to 
experimental results.  In all cases we obtained a mixing angle
consistent with 
$\theta_K \simeq 45^o$.  There are two important conclusions we can 
draw from this result.  First, the relativized quark model predicts 
a much smaller mixing angle of $\sim -5^o$.  Either the quark model 
result is way off, which is possible given the delicate cancellation 
taking place between the contributions to the spin-orbit term,
or a different mechanism is responsible for the 
$^3P_1-^1P_1$ mixing \cite{lipkin77}.  
The second observation we make 
on the basis of the quark model results is that the heavy quark 
limit does not appear to be applicable to the strange axial mesons.  
We come to this conclusion because the tensor interaction is still 
comparable in size to the spin-orbit interactions and additionally, the 
mixing angle is not compatible with that expected in the 
heavy quark limit.

\acknowledgments

This research was supported in part by the Natural Sciences and
Engineering Research Council of Canada. 
The authors thank Sherry Towers for discussion and 
S.G. thanks Nathan Isgur for helpful conversations.

\begin{figure}
\centerline{\epsfig{file=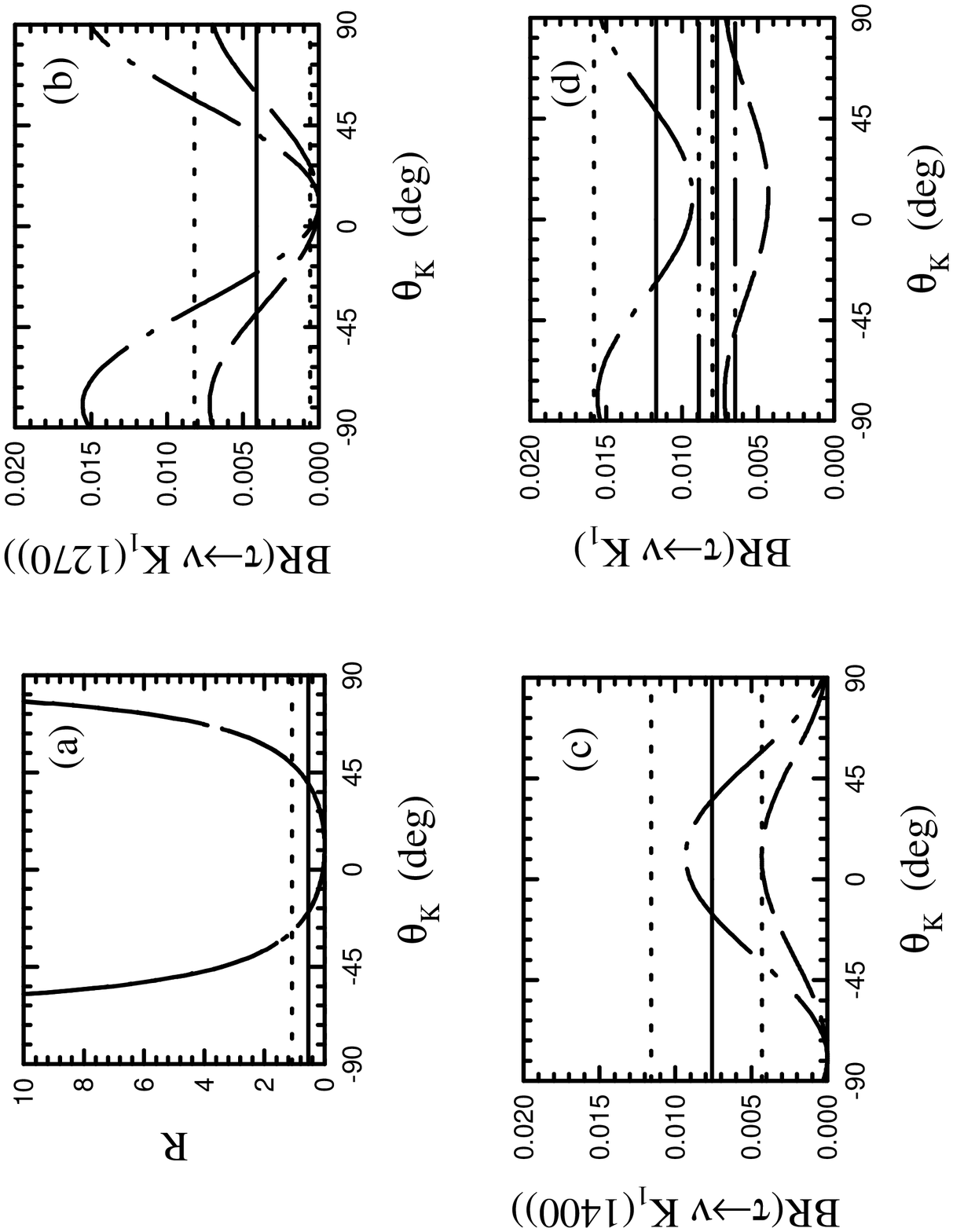,width=12.0cm,clip=}}
\caption[]{The $\tau \to K_1 \nu$ decay widths as a function of 
$\theta_K$ for the non-relativistic results.  
$R=BR (\tau \to \nu K_1 (1270))/BR (\tau \to \nu K_1 
(1400))$. In all figures the dashed curve is for the wavefunctions 
from Ref.~\cite{colic83} and the dot-dashed curve is for the wavefunctions 
from Ref.~\cite{isgw89}.  In the figure for $R$ both curves lie on each 
other.  The solid and dotted lines are for the experimental values 
and their $1-\sigma$ errors from the TPC/Two-Gamma 
measurement \cite{tpc95}.  In addition, in the figure for 
$BR(\tau\to \nu K_1)$, the solid line bounded by the 
dot-dot-dashed lines are for the CLEO/ALEPH result and their $1-\sigma$ error
\cite{alemany94}.}
\end{figure}

\newpage

\begin{figure}
\centerline{\epsfig{file=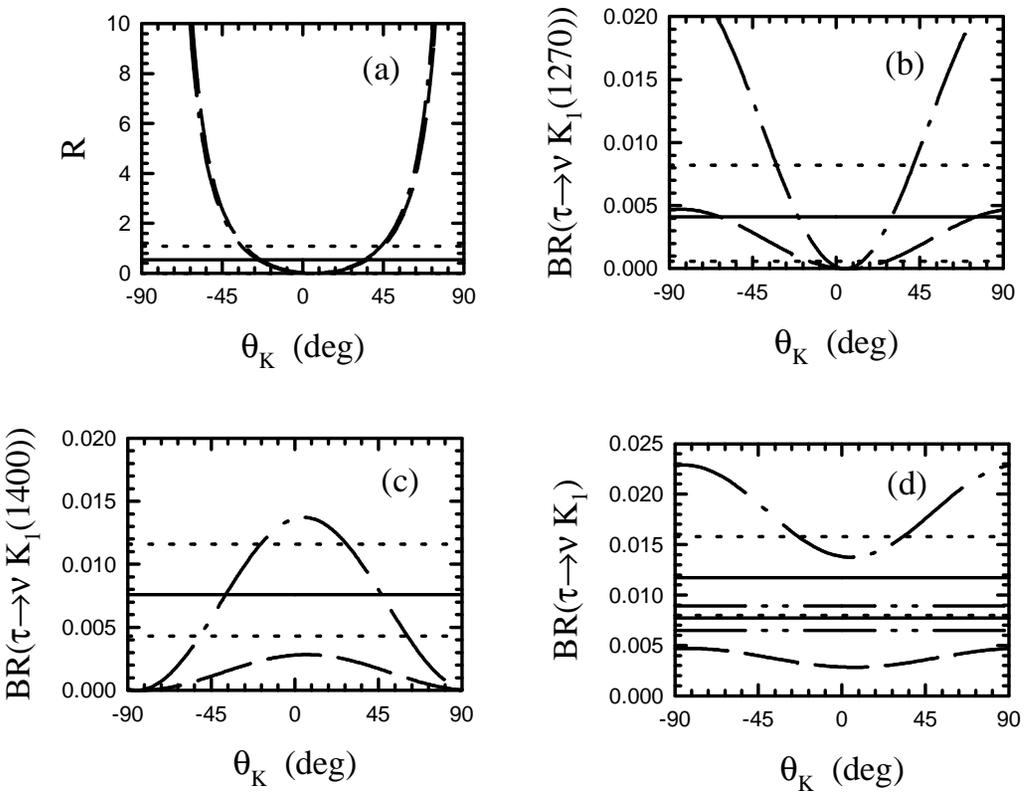,width=12.0cm,clip=}}
\caption{The $\tau \to K_1 \nu$ decay widths as a function of 
$\theta_K$ for the relativized results.  The line labelling is the 
same as in Fig.~1.}
\end{figure}

\newpage

\begin{figure}
\centerline{\epsfig{file=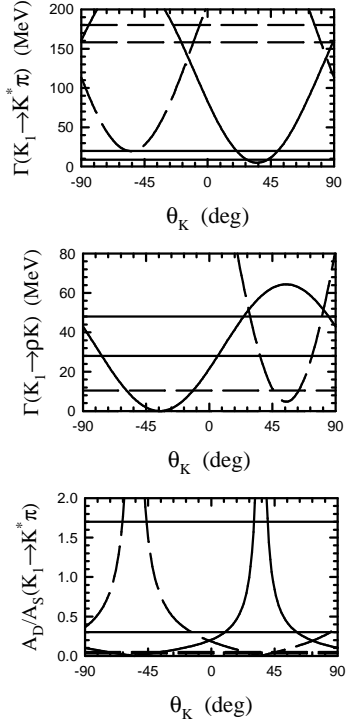,width=8.0cm,clip=}}
\caption[]{Predictions of the pseudoscalar emission model as a function of 
$\theta_K$ for the $K_1$ partial widths (to $K^* \pi$ and $\rho K$), and ratio 
of D to S amplitudes (to $K^* \pi$).  The solid curves are for the 
$K_1(1270)$ and the dashed curves are for the $K_1(1400)$.
The horizontal lines are the $1-\sigma$ 
error bounds for the experimental measurements \cite{pdb} with the same line 
labelling (solid, dashed) as the predictions. 
(The experimental lower bound for 
$K_1(1400)\to \rho K$ lies on the axis.)}
\end{figure}

\newpage

\begin{figure}
\centerline{\epsfig{file=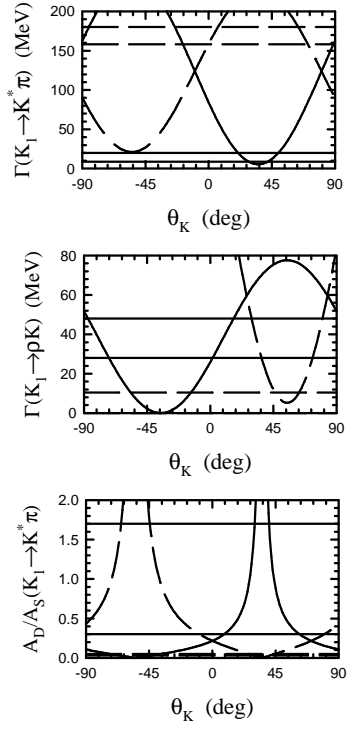,width=8.0cm,clip=}}
\caption{Predictions of the $^3P_0$ model as a function of 
$\theta_K$ for the $K_1$ partial widths (to $K^* \pi$ and $\rho K$), and ratio 
of D to S amplitudes (to $K^* \pi$).  The line labelling is the same as in 
Fig.~3.}
\end{figure}

\begin{table}
\caption{Axial vector decay constants in the nonrelativistic limit 
using simple harmonic oscillator wavefunctions. The $f_{K_1}$ for 
$|K^+ \rangle = -|u\bar{s}\rangle$ are given in 
units of GeV$^2$.}
\label{table1}
\begin{tabular}{lll}
Parameter set & $M_{K_1}=1.273$~GeV & $M_{K_1}=1.402$~GeV \\
\tableline
Set 1$^a$ & $f_{K_b} = -0.049$ &  $f_{K_b} = -0.050$\\
	 & $f_{K_a} = +0.269$ &  $f_{K_a} = +0.283$\\
Set 2$^b$ & $f_{K_b} = -0.070$ &  $f_{K_b} = -0.073$\\
	 & $f_{K_a} = +0.396$ &  $f_{K_a} = +0.415$\\
\end{tabular}
$^a$ From Ref. \cite{colic83} with $\beta =0.257$~GeV, 
$m_u=m_d=0.33$~GeV and $m_s=0.55$~GeV. \\
$^b$ From Ref. \cite{isgw89} with $\beta=0.3$ and $m_u=m_d$ and $m_s$ 
as above.
\end{table}

\begin{table}
\caption{Axial vector decay constants using the relativized 
mock-meson matrix elements.  Simple harmonic oscillator wavefunctions are 
used with the parameters given below. The $f_{K_1}$ for 
$|K^+ \rangle = -|u\bar{s}\rangle$are given in 
units of GeV$^2$.}
\label{table2}
\begin{tabular}{lll}
Parameter set & $M_{K_1}=1.273$~GeV & $M_{K_1}=1.402$~GeV \\
\tableline
Set 1$^a$ & $f_{K_b} = -0.024$ &  $f_{K_b} = -0.025$\\
	 & $f_{K_a} = +0.220$ &  $f_{K_a} = +0.231$\\
Set 2$^b$ & $f_{K_b} = -0.040$ &  $f_{K_b} = -0.042$\\
	 & $f_{K_a} = +0.486$ &  $f_{K_a} = +0.510$\\
\end{tabular}
$^a$ From Ref. \cite{isgw89} with $\beta=0.3$, 
$m_u=m_d=0.33$~GeV and $m_s=0.55$~GeV. \\
$^b$ We used effective oscillator parameters from Ref. \cite{godfrey85}.  
They were obtained by fitting simple harmonic oscillator wavefunctions to 
the rms radii of the wavefunctions of Ref.~\cite{godfrey85} to 
obtain $\beta_{K_1}=0.45$~GeV.  $m_u=m_d=0.22$~GeV and $m_s=0.419$~GeV. 
\end{table}

\begin{table}
\caption{Strong decay amplitudes for the strange axial mesons using 
the pseudoscalar emission model, the $^3P_0$ decay model and the 
flux-tube breaking model.  
Note that the amplitudes include phase space and are all given in 
units of MeV$^{1/2}$.}
\label{table3}
\begin{tabular}{lcccc}
Amplitude & Pseudoscalar & $^3P_0$ & \multicolumn{2}{c}{Flux-tube breaking}\\
	& emission & Set 1$^a$ & Set 1$^a$ & Set 2$^b$\\
$\gamma$  &	& 6.25 & 10.4 & 12.8 \\
\tableline
$S(K^{low}_1 \to \rho K)$ & 8.02 & 8.81 & 8.50 & 11.0 \\
$D(K^{low}_1 \to \rho K)$ & 0.074 & 0.056 & 0.057 & 0.055 \\
$S(K^{low}_1 \to K^* \pi)$ & 15.5 & 15.5 & 14.8 & 20.5 \\
$D(K^{low}_1 \to K^* \pi)$ & 2.23 & 2.36 & 2.43 & 2.29 \\
$S(K^{high}_1 \to \rho K)$ & 15.4 & 15.5 & 14.8 & 20.1 \\
$D(K^{high}_1 \to \rho K)$ & 2.18 & 2.28 & 2.34 & 2.25 \\
$S(K^{high}_1 \to K^* \pi)$ & 17.3 & 15.1 & 14.2 & 20.7 \\
$D(K^{high}_1 \to K^* \pi)$ & 4.45 & 4.61 & 4.74 & 4.44 \\
$S(K^{high}_1 \to K \omega)$ & 15.1 & 15.4 & 14.7 & 20.0 \\
$D(K^{high}_1 \to K \omega)$ & 1.96 & 2.04 & 2.10 & 2.02 \\
\end{tabular}
$^a$ Simple harmonic oscillator wavefunctions with
$\beta=0.40$~GeV, $m_u=0.33$~GeV, and $m_s=0.55$~GeV. \\
$^b$ Wavefunctions from Ref. \cite{godfrey85}.
\end{table}

\begin{table}
\caption{Partial decay widths and ratios of D to S amplitudes of the 
strange axial mesons for 
the pseudoscalar emission model, the $^3P_0$ decay model and the 
flux-tube breaking model using the fitted value of $\theta_K$. The 
widths are given in MeV. $A_D/A_S$ refers to the ratio of $D$ to $S$ 
amplitudes. The errors on $\theta_K$ are $1-\sigma$.}
\label{table4}
\begin{tabular}{llllll}
Decay & Experiment &
	 Pseudoscalar & $^3P_0$ & \multicolumn{2}{c}{Flux-tube breaking$^a$}\\
	& (RPP) 
	& emission & Set 1$^b$ & Set 1$^b$ & Set 2$^c$\\
\tableline
$\theta_K$ &  & $48^o \pm 5^o$ & $ 45^o \pm 4^o$ & $44^o\pm 4^o$ & 
	$51^o \pm 3^o$ \\
\tableline
$\Gamma(K_1 (1270) \to \rho K)$ & $38 \pm 10$
	 & 63 & 75 & 70 & 121 \\
$\Gamma(K_1 (1270) \to K^* \pi)$ & $14.4 \pm 5.5$
	 & 16 & 12 & 11 & 35 \\
$|A_D/A_S (K_1 (1270) \to K^* \pi)|$ & $1.0\pm 0.7$
	 & 0.64 & 0.89 & 1.02 & 0.40 \\
$\Gamma(K_1 (1400) \to \rho K)$ & $5.2 \pm 5.2$
	 & 7.9 & 12 & 13 & 6.7 \\
$\Gamma(K_1 (1400) \to K^* \pi)$ & $164\pm 16$
	 & 286 & 221 & 197 & 400 \\
$\Gamma(K_1 (1400) \to K \omega)$ & $1.7\pm 1.7$
	& 2.3 & 3.6 & 3.9 & 1.9 \\
$|A_D/A_S(K_1 (1400) \to K^* \pi)|$ & $0.04 \pm 0.01$
	& 0.058 & 0.052 & 0.051 & 0.062 
\end{tabular}

$^a$ Note that because the flux-tube breaking calculation involves a
numerical integral with a 1\% error, the two values of $S$ (or $D$)
calculated from the $K_1(1270)$ and $K_1(1400)$ decay results may not
agree exactly.  In Table III an average value of the two results 
is given.  Because the
$S$ and $D$ values in Table III are not exact, using them with
Eqn.~\ref{sddef} will not exactly reproduce the flux-tube breaking
results show in this table, which are calculated directly from the
numerical work. \\
$^b$ Simple harmonic oscillator wavefunctions with
$\beta=0.40$~GeV, $m_u=0.33$~GeV, and $m_s=0.55$~GeV. \\
$^c$ Wavefunctions from Ref. \cite{godfrey85}.
\end{table}

\end{document}